\documentclass[10pt,twocolumn]{article}
\usepackage{graphics}
\usepackage{amssymb}
\title{\Large\bf Strong Local Passivity in Finite Quantum Systems}
\author{Michael Frey$^1$, Ken Funo$^2$, and Masahiro Hotta$^3$   \\[2mm]
{\it\normalsize $^1$Department of Mathematics, Bucknell University, Lewisburg, PA 17837 USA}     \\
{\it\normalsize $^2$Department of Physics, The University of Tokyo, Tokyo 113-0033, Japan}      \\
{\it\normalsize $^3$Graduate School of Science, Tohoku University, Sendai 980-8578, Japan}  \\[1mm]  }

\date{
\parbox{6.4in}{\normalsize
\vspace*{3mm}
Passive states of quantum systems are states from which no system energy can be extracted by any cyclic
(unitary) process. Gibbs states of all temperatures are passive. Strong local (SL) passive states are
defined to allow any general quantum operation, but the operation is required to be local, being applied
only to a specific subsystem. Any mixture of eigenstates in a system-dependent neighborhood of a
nondegenerate, entangled ground state is found to be SL passive. In particular, Gibbs states are SL
passive with respect to a subsystem only at or below a critical, system-dependent temperature. SL
passivity is associated in many-body systems with the presence of ground state entanglement in a way
suggestive of collective quantum phenomena such as quantum phase transitions, superconductivity, and
the quantum Hall effect. The presence of SL passivity is detailed for some simple spin systems where
it is found that SL passivity is neither confined to systems of just a few particles nor limited to
the near vicinity of the ground state.
}
}

\begin{document}
\maketitle

\begin{center}
{\bf I. INTRODUCTION}
\end{center}
\vspace*{-1mm}

The maximum energy that can be extracted from a physical system---the work that it can
do---by an applied process is a fundamental thermodynamic problem of continuing interest
\cite{alic,funo}. This problem is typically posed \cite{pusz,lena,tasa} for a quantum system
with Hamiltonian $\mbox{\bf H}$ in terms of cyclic (unitary) processes in which the system,
initially isolated, is coupled at time $t=0$ to external sources of work with combined potential
$\mbox{\bf V}(t)$ and later decoupled from them at time $t=\tau$, creating for $t\in [0,\tau]$
a time-dependent system Hamiltonian $\mbox{\bf H}(t) = \mbox{\bf H} + \mbox{\bf V}(t)$ (with
associated free energy $F(t)=-kT\ln\mbox{Tr}[\exp(-\mbox{\bf H}(t)/kT)]$). System states for
which no cyclic process can extract a positive amount of energy from the system are called
passive \cite{pusz,tasa}. More specifically, since in this context the change in free energy
is zero, a system in a passive state can do no positive work. An important result for finite
quantum systems is that Gibbs states are passive. Indeed, this is the {\it no perpetuum mobile} version
of the second law of thermodynamics for equilibrium as formulated by Thomson \cite{terl,alla}.

We introduce in this paper a variant of passivity we call strong local (SL) passivity, which
identifies a new collective quantum phenomenon exhibited by multipartite systems. We find for
finite quantum systems with a nondegenerate, entangled ground state that states in a neighborhood
of the ground state are SL passive. In particular, though all Gibbs states (of any temperature)
are passive, only Gibbs states at or below a critical, system-dependent temperature are SL passive.
This means for many-body systems that, for any state close to the ground state, the
ground state entanglement and nondegeneracy sufficiently constrain the system's subsystems to
collectively inhibit energy release from any subsystem. Ground state entanglement is a recognized
root cause of other collective quantum phenomena, including quantum phase transitions \cite{sach},
superconductivity \cite{laug}, and the quantum Hall effect \cite{bard}.

A system state is defined to be SL passive if no general (Kraus, operator-sum) quantum operation
${\cal G}$ applied locally to a subsystem can extract positive energy from the system. We are
restricted, in other words, to operations of the form ${\cal G} \otimes {\cal I}$ where ${\cal I}$
is the identity operation for the rest of the system. The system dynamics driven by $\mbox{\bf H}$
can have, generally, a nonlocal component. So that the effect of ${\cal G}$ applied locally is not
confounded with the time evolution accompanying any nonlocal component of $\mbox{\bf H}$,
we include in SL passivity's definition the idealization that $\cal G$ proceeds much faster than
the system's natural unitary evolution $\exp(-i\mbox{\bf H}\tau/\hbar)$ due to $\mbox{\bf H}$.
In fact, fast local operations are of main interest in applications; in, for example, circuit-based
quantum information processing, gates must operate faster than the background evolution of
the physical substrate. For sufficiently fast ${\cal G}$ and the system in a state $\rho$, the
energy extracted by $\cal G$ is effectively
\begin{equation}
\Delta E(\rho) = \mbox{Tr}[ \mbox{\bf H}\rho ]-\mbox{Tr}\left[\mbox{\bf H}({\cal G}\otimes {\cal I})(\rho)\right] .
\label{eq:sysen}
\end{equation}
\noindent
The local energy $\Omega_\circ$ of a subsystem is defined to be the maximum of $\Delta E(\rho)$
for any $\cal G$ \cite{frey}. Note that $\Omega_\circ\geq 0$ and that $\rho$ is SL passive
if and only if $\Omega_\circ=0$. Local energy for SL passive states is analogous to
ergotropy introduced for state passivity \cite{all2}.

Our definition of SL passivity reflects two modifications of the usual notion of state passivity.
First, any general quantum operation ${\cal G}$ expressible in terms of Kraus operators
\cite{niel} is allowed, relaxing the restriction to unitary operations. Second, only a subsystem
is accessible, making ${\cal G}$ local to just that subsystem. It is easy to check that neither
modification alone yields interesting physics. Suppose we define SL passivity to allow any general
quantum operation $\cal G$ but do not narrow the operation's scope to a subsystem. For any
finite quantum system with a ground state $|E_0\rangle$ and eigenstates $|E_k\rangle$ of higher
energy, a $\cal G$ can be constructed from Kraus operators $\mbox{\bf K}_k = |E_0\rangle\langle E_k |$
so that ${\cal G} (\rho) = |E_0\rangle\langle E_0|$ for any system state $\rho$. By this definition,
only a ground state can be SL passive. Or, suppose we narrow the scope of the
operation to a subsystem but still require a unitary operation; that is, we allow only operations
of the form ${\cal U} \otimes {\cal I}$ where ${\cal U}$ is a local unitary operation on the subsystem.
Here again nothing results; ${\cal U} \otimes {\cal I}$ is unitary, so for systems with identifiable
subsystems all passive states, including Gibbs states, would be SL passive. Only together do the
two modifications have an unexpected and interesting result.

Conditions for SL passivity are presented as a theorem in the following section. Then in
section III the presence of SL passivity is detailed in some simple quantum spin systems.
Section IV addresses the possible extent of SL passivity in a system in terms of type of
state, number of system particles, and size of the ground state's SL passivity
neigborhood. We close in the last section with some summary remarks.

\begin{center}
{\bf II. MAIN RESULT}
\end{center}
\vspace*{-1mm}

The setting of our main result for SL passivity is a finite quantum system ${\cal S}$ (described
by a complex Hilbert space ${\cal H}$ with $d=\mbox{dim}({\cal H})<\infty$) with a subsystem
(or component) $\cal C$ whereby ${\cal H} = {\cal H}_c \otimes {\cal H}_{\bar{c}}$ where ${\cal H}_c$
and ${\cal H}_r$ are the Hilbert spaces associated with $\cal C$ and the rest of $\cal S$,
respectively. The Hamiltonian of $\cal S$ is
\begin{equation}
\mbox{\bf H} = \sum_{k=0}^{d-1} E_k |E_k \rangle\langle E_k|
\label{eq:spec}
\end{equation}
\noindent
with eigenstates $|E_k \rangle$ and associated eigenenergies $E_0 \leq E_1 \leq \ldots \leq E_{d-1}$.
The Schmidt decomposition of the ground state $|E_0\rangle$ is \cite{niel}
\begin{equation}
|E_0\rangle = \sum_s \sqrt{q_s} |c_s\rangle|r_s\rangle
\label{eq:schdec}
\end{equation}
\noindent
where $\sum_s q_s=1$ and $|c_s\rangle$ and $|r_s\rangle$ are, respectively, orthonormal states
of ${\cal H}_c$ and ${\cal H}_r$. The ground state $|E_0\rangle$ is fully entangled if all the $q_s$
in (\ref{eq:schdec}) are positive \cite{niel,brus}.

We will be concerned mostly with system states $\rho$ of $\cal S$ that commute with $\mbox{\bf H}$;
in other words, eigenmixtures
\begin{equation}
\rho = \sum_{k=0}^{d-1} p_k |E_k\rangle\langle E_k |
\label{eq:stmix}
\end{equation}
\noindent
that are statistical mixtures of the eigenstates $|E_k\rangle$ with population
probabilities $p_k$ such that $\sum_k p_k =1$. Eigenmixtures (\ref{eq:stmix}) disallow coherences
among the system eigenstates, but still include the important case of Gibbs states for which
\begin{equation}
p_k = \frac{1}{\cal Z} \exp\left( -\frac{E_k}{kT} \right)
\label{eq:gib}
\end{equation}
\noindent
where $k$ is Boltzmann's constant, $T$ is Gibbs temperature, and
${\cal Z}=\mbox{Tr}[\exp(-\mbox{\bf H}/ kT)]$ is the partition function. Eigenmixtures play a
distinctive role in connection with system passivity; for example \cite{lena}, a state of a finite
quantum system is passive if and only if it is an eigenmixture with $p_k \geq p_{k^\prime}$
for $E_k < E_{k^\prime}$. We will see that with respect to SL passivity eigenmixtures play a
similarly prominent role. We now state our main result.

{\it Theorem}: Let $\cal S$ be a finite quantum system with Hamiltonian (\ref{eq:spec}) and a
specified subsystem $\cal C$ such that $\cal C$ is fully entangled with the rest of $\cal S$
in the ground state $|E_0\rangle$. Suppose further that $|E_0\rangle$ is nondegenerate; that
is, $E_0 < E_1$. Then a threshold ground state population probability $p_*<1$ exists such that
any eigenmixture $\rho$ in (\ref{eq:stmix}) with $p_0 \geq p_*$ is SL passive.

{\it Proof}: Let $\cal G$ be a general quantum operation \cite{niel} on subsystem $\cal C$.
For states $\sigma$ on ${\cal H}_c$,
\begin{equation}
{\cal G}(\sigma) = \sum_{\mu} \mbox{\bf K}_{\mu} \,\sigma\, \mbox{\bf K}_\mu^{\dagger}
\label{eq:gqop}
\end{equation}
\noindent
with Kraus operators $\mbox{\bf K}_\mu$ on ${\cal H}_c$ such that
\begin{equation}
\sum_\mu  \mbox{\bf K}_\mu^{\dagger} \mbox{\bf K}_\mu = \mbox{\bf I} \, .
\label{eq:kid}
\end{equation}
\noindent
With $\cal S$ initially
in the eigenstate $|E_k\rangle$, the system energy loss due to ${\cal G}$ is
\begin{displaymath}
\Delta E_k = E_k - \mbox{Tr}\left[ \mbox{\bf H} ({\cal G}\otimes {\cal I})(|E_k\rangle\langle E_k|) \right] \, .
\end{displaymath}
\noindent
A calculation involving (\ref{eq:spec}), (\ref{eq:kid}) and the completeness identity
$\sum_k |E_k\rangle\langle E_k | = \mbox{\bf I}$ then yields
\begin{equation}
\Delta E_k = \sum_{k^\prime \neq k} (E_k - E_{k^\prime})\sum_\mu \left| \langle E_{k^\prime}
|\mbox{\bf K}_\mu | E_k\rangle \right|^2     \, .
\label{eq:deriv}
\end{equation}
\noindent
Because $|E_0\rangle$ is nondegenerate, $\Delta E_0\leq 0$ in (\ref{eq:deriv}) and
\begin{eqnarray}
\Delta E_0 = 0 \!\!\!\!\!\!\!\!\!\!
&& \Leftrightarrow   \langle E_m |\mbox{\bf K}_\mu | E_0\rangle =0 \; \forall m\neq 0 \label{eq:nondegen} \\
&& \Leftrightarrow \mbox{\bf K}_\mu | E_0\rangle = \lambda_\mu | E_0\rangle               \nonumber \\
&& \Leftrightarrow \sum_s \sqrt{q_s} ( \mbox{\bf K}_\mu |s\rangle_c - \lambda_\mu |s\rangle_c ) |s\rangle_r = 0
\label{eq:schmidt} \\
&& \Leftrightarrow  \mbox{\bf K}_\mu = \lambda_\mu  \mbox{\bf I}_c  \label{eq:trivial}
\end{eqnarray}
\noindent
where (\ref{eq:nondegen}) holds because $|E_0\rangle$ is nondegenerate, (\ref{eq:schmidt}) follows
from (\ref{eq:schdec}), and (\ref{eq:trivial}) is due to $q_s \neq 0$ for all $s$. Now consider
$\mbox{\bf K}_\mu$ in a neigborhood of the trivial operator $\lambda_\mu  \mbox{\bf I}_c$.
The Kraus operators $\mbox{\bf K}_\mu$ are trace-class (hence compact) acting on the
finite-dimensional Hilbert space ${\cal H}_c$, so
\begin{equation}
\mbox{\bf K}_\mu = \lambda_\mu \mbox{\bf I} +
\sum_\gamma \theta_\gamma \mbox{\bf J}_{\mu\gamma} +
\frac{1}{2} \sum_{\gamma ,\gamma^\prime} \theta_\gamma \theta_{\gamma^\prime}
\mbox{\bf J}_{\mu\gamma\gamma^\prime}
\label{eq:exp}
\end{equation}
\noindent
for small $\theta_\gamma > 0$ to order
$O(\theta_\gamma \theta_{\gamma^\prime} \theta_{\gamma^{\prime\prime}})$.
Put (\ref{eq:exp}) into (\ref{eq:deriv}), with
$\chi_{\mu\gamma}=\langle E_{k^\prime}|\mbox{\bf J}_{\mu\gamma}|E_k\rangle$. Then,
using $\langle E_{k^\prime} |\lambda_\mu \mbox{\bf I} |E_k\rangle = 0$ for $k^\prime \neq k$, we have
\begin{equation}
\Delta E_k = \sum_{\gamma,\gamma^{\prime}} \theta_\gamma \theta_{\gamma^\prime}
\sum_{k^\prime \neq k} (E_k - E_{k^\prime})\sum_\mu
\chi_{\mu\gamma}^\dagger \chi_{\mu \gamma^\prime}
\label{eq:scd}
\end{equation}
to order $O(\theta_\gamma \theta_{\gamma^\prime} \theta_{\gamma^{\prime\prime}})$ for each
$k\geq 1$. The remarkable feature of (\ref{eq:scd}), and the key to the proof, is that no term
linear in $\theta_\gamma$ appears for $k\geq 1$. (Linear terms do appear when we attempt to
adjust the proof for states that are not eigenmixtures.)
The absence of linear terms in (\ref{eq:scd}) means that $\Delta E_k/\Delta E_0$ does
not diverge for $\theta_\gamma \rightarrow 0$. So, for any $k\geq 1$ and any nontrivial ${\cal G}$,
\begin{displaymath}
\frac{1-p_0}{p_0}\left| \frac{\Delta E_k}{\Delta E_0}\right| \leq 1
\end{displaymath}
\noindent
for all $p_0 \geq p_*$ for some $p_*<1$. Since $\Delta E_0<0$ for all nontrivial ${\cal G}$,
\begin{displaymath}
\frac{p_k}{1-p_0} p_0\Delta E_0 + p_k|\Delta E_k| \leq 0  \, ,
\end{displaymath}
\noindent
from which, by summing, follows
\begin{equation}
p_0\Delta E_0 + \sum_{k=1}^{d-1} p_k |\Delta E_k | \leq 0
\label{eq:key}
\end{equation}
\noindent
for all $p_0 \geq p_*$. If the state $\rho$ of $\cal S$ is an eigenmixture (\ref{eq:stmix}), the system energy
$\Delta E(\rho)$ in (\ref{eq:sysen}) extracted by the local operation ${\cal G}$ is
$\Delta E(\rho) = \sum_{k=0}^{d-1} p_k \Delta E_k$, and we conclude from (\ref{eq:key}) that,
for any eigenmixture $\rho$,
$\Delta E(\rho) \leq 0$ for all $p_0 \geq p_*$.      \hfill $\Box$

{\it Corollary}: The Gibbs states of a finite quantum system with nondegenerate, fully entangled
ground state are SL passive with respect to a subsystem for all temperatures $T \leq T_*$ for some
critical temperature $T_* > 0$.

Our theorem is stated for a system with just some identified subsystem. For a many-particle system
governed by, say, a particle-symmetric Hamiltonian, SL passivity with respect to one particle implies
SL passivity for all, and the theorem then says that the system's particles in a eigenmixture
sufficiently near $|E_0\rangle$ are constrained by $|E_0\rangle$'s entanglement and nondegeneracy
to collectively disallow energy release from {\it any} particle.

\vspace*{1mm}
\begin{center}
{\bf III. TWO-PARTICLE SYSTEMS}
\end{center}
\vspace*{-1mm}

Our theorem and its corollary can be seen at work in a variety of multi-particle quantum systems. We
detail this in this section in examples of two-particle systems.

Let ${\cal S}_2$ be a pair of coupled spin-$\frac{1}{2}$ particles with Hamiltonian
\begin{equation}
\mbox{\bf H} = \kappa \mbox{\boldmath$\sigma$}_1^x  \mbox{\boldmath$\sigma$}_2^x
+ \mbox{\boldmath$\sigma$}_1^z
+ \mbox{\boldmath$\sigma$}_2^z
\label{eq:ham2}
\end{equation}
\noindent
where the Pauli operator terms $\mbox{\boldmath$\sigma$}_1^z$ and $\mbox{\boldmath$\sigma$}_2^z$
reflect the presence of an external magnetic field transverse to the coupling and $\kappa>0$ is the
coupling's relative strength. The pair ${\cal S}_2$ has a fully entangled ground state and eigenenergies
\begin{equation}
E_0 = -m \, , \;\,   E_1 = -\kappa \, , \;\, E_2 = \kappa \, , \;\,   E_3 = m
\label{eq:en}
\end{equation}
\noindent
where $m=\sqrt{\kappa^2+4}$.

Consider a general quantum operation $\cal G$ of the form (\ref{eq:gqop}) applied to a particle of
${\cal S}_2$. For a spin-$\frac{1}{2}$ particle, $\cal G$ requires at most four Kraus operators \cite{niel}
\begin{equation}
\mbox{\bf K}_\mu = \left(  \matrix{ s_\mu & t_\mu \cr u_\mu & v_\mu }  \right)
\label{eq:ko}
\end{equation}
\noindent
with complex-valued elements $s_\mu, t_\mu, u_\mu, v_\mu$. In terms of these elements,
condition (\ref{eq:kid}) for the $\mbox{\bf K}_\mu$ becomes
\begin{eqnarray}
&& \!\!\!\!\!\!\!\!\!\!
\mbox{\bf s}^\dagger \mbox{\bf s} + \mbox{\bf u}^\dagger \mbox{\bf u} = 1  \nonumber \\
&& \!\!\!\!\!\!\!\!\!\!
\mbox{\bf t}^\dagger \mbox{\bf t} + \mbox{\bf v}^\dagger \mbox{\bf v} = 1  \label{eq:krc} \\
&& \!\!\!\!\!\!\!\!\!\!
\mbox{\bf s}^\dagger \mbox{\bf t} + \mbox{\bf u}^\dagger \mbox{\bf v} = 0  \nonumber
\end{eqnarray}
\noindent
where $\mbox{\bf s}=(s_1 \; s_2 \; s_3 \; s_4)^\top$, etc.

\begin{figure}[b!]
\centerline{\scalebox{1}{\includegraphics{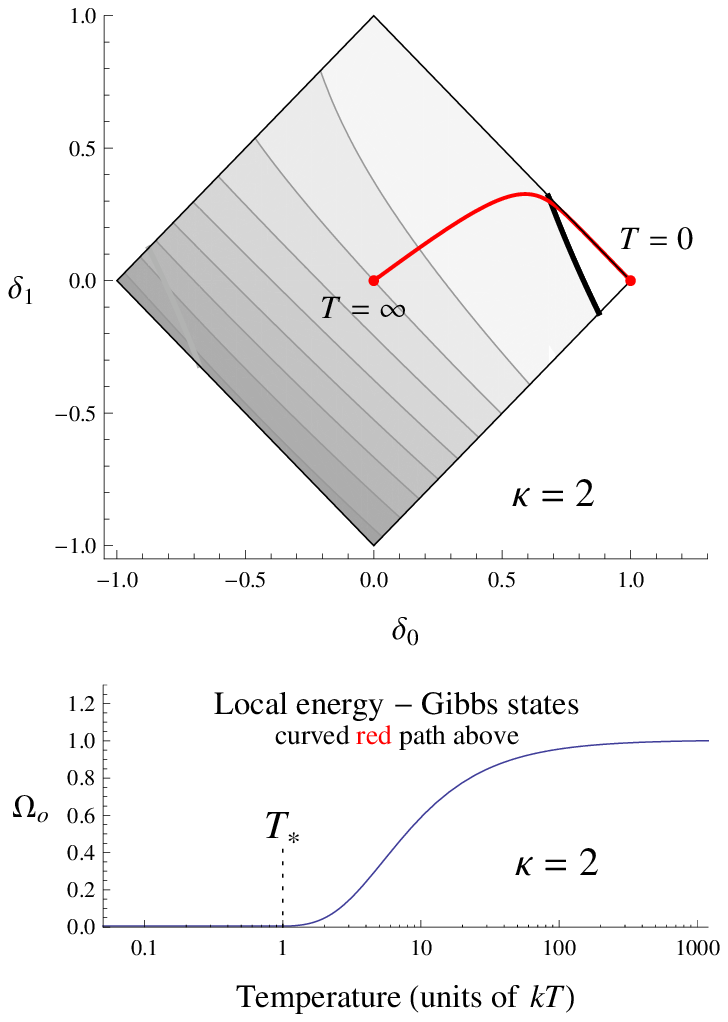}}}
\vspace*{2mm}
\begin{center}
\parbox{2.9in}{\small
FIG.\ 1. Local energy $\Omega_\circ$ for $\kappa=2$ in $\cal S$ with system state $\rho$
parameterized by the probability differences $\delta_0$, $\delta_1$. Darker shading indicates
greater local energy. $\Omega_\circ = 0$ for any eigenmixture $\rho$ in a neighborhood of
the ground state $|E_0\rangle$. Gibbs states are SL passive below the critical temperature $T_*$.}
\end{center}
\end{figure}

The local energy $\Omega_\circ$ in (\ref{eq:maxxx}) depends solely on $\eta$ and $\xi$ in (\ref{eq:etxi}),
which depend in turn on the eigenmixture $\rho$ through just the two probability differences $\delta_0$
and $\delta_1$. Because of this limited dependence on $\rho$, an eigenmixture can be SL passive
without being passive. For $\kappa=2$ in ${\cal S}_2$, for example, the eigenmixture
$(p_0,p_1,p_2,p_3) = (.96,0,.04,0)$ is SL passive ($\Omega_\circ=0$), but not passive ($p_2 > p_1$).
In general, passivity is neither necessary nor sufficient for SL passivity.

For the pair ${\cal S}_2$ in an eigenmixture state $\rho$, the energy extracted by applying
$\cal G$ locally to a particle in ${\cal S}_2$ is, after elementary calculation,
\begin{eqnarray}
&& \!\!\!\!\!\! \!\!\!\!\!\! \!\!\!\!\!\! \Delta E(\rho) = (1-\eta) \mbox{\bf u}^\dagger \mbox{\bf u}
- (1+\eta)\mbox{\bf t}^\dagger \mbox{\bf t}                  \nonumber       \\
&& \;\;\;\;\; + \, \xi\frac{\mbox{\bf s}^\dagger \mbox{\bf v} + {\mathbf v}^\dagger \mbox{\bf s}
+ \mbox{\bf u}^\dagger \mbox{\bf t}  + \mbox{\bf t}^\dagger \mbox{\bf u}}{2} - \xi    \label{eq:Ome}
\end{eqnarray}
\noindent
where
\begin{equation}
\eta = \frac{2}{m}\delta_0 \; , \;\;\;\;\;
\xi = \frac{\kappa^2}{m}\delta_0 + \kappa\delta_1
\label{eq:etxi}
\end{equation}
\noindent
with $\delta_0=p_0-p_3$ and $\delta_1=p_1-p_2$. The maximum of (\ref{eq:Ome}) subject to (\ref{eq:krc})
is the local energy $\Omega_\circ$ of a particle in ${\cal S}_2$. We twice apply the dominance
argument in \cite{frey}, once for $\xi\geq 0$ and then again for $\xi<0$. We find that the local
energy in a particle of ${\cal S}_2$ is $\Omega_\circ =\Omega_\circ (\eta,\xi)$ where
\begin{equation}
\Omega_\circ (\eta,\xi) =
\cases{ \sqrt{\frac{1-\eta^2+\xi^2}{1-\eta^2}} -\xi - \eta \, , & $\!\!\! |\eta \xi| <1-\eta^2$ \\[2mm] \cr
|\xi|+|\eta| -\xi - \eta   \, , & $\!\!\!\!$ otherwise }   .
\label{eq:maxxx}
\end{equation}
\noindent
Shown in the top display of Fig.\ 1 is a contour plot of $\Omega_\circ$ for $\kappa=2$. The plot's
diamond-shaped domain given by $|\delta_0|+|\delta_1|\leq 1$ is all possible combinations of $\delta_0$
and $\delta_1$. Of special interest in the $\delta_0,\delta_1$ parameter space is the three-sided
region at right that includes $\delta_0=1$. This region consists of the ground state $|E_0\rangle$
and all eigenmixtures $\rho$ that are small departures from it. These states are
all SL passive; they all have $\eta,\xi \geq 0$ with zero local energy,
$\Omega_{\mbox{\tiny o}}=|\xi|+|\eta| -\xi - \eta =0$. The extent of the
SL passive neighborhood around $|E_0\rangle$ can be quantified by considering the
three-sided $\Omega_\circ=0$ region in Fig.\ 1's top display. A sufficient condition for
$\Omega_\circ=0$ is that the system eigenmixture have $\delta_0 \geq \delta_\ast$ where
\begin{displaymath}
\delta_\ast = \frac{\kappa+\sqrt{(3m^2+2\kappa m-8)}}{2(m^2+\kappa m-2)}
\end{displaymath}
\noindent
is the $\delta_0$ coordinate of the bottom corner of the $\Omega_{\mbox{\tiny o}}=0$ region
(see top display in Fig.\ 1). A sufficient condition for $\delta_0 \geq \delta_\ast$ is, in turn,
\begin{equation}
p_0 \geq p_\ast = \frac{1+\delta_\ast}{2} \, .
\label{eq:thrprob}
\end{equation}
\noindent
Any eigenmixture $\rho$ of the form (\ref{eq:stmix}) with ground state probability $p_0 \geq p_\ast$ has
zero local energy. The threshold probability $p_\ast$ in (\ref{eq:thrprob}) is a decreasing function of $\kappa$
with, for example, $p_\ast = .9383$ for $\kappa=2$. We conclude that the neighborhood of $|E_0\rangle$
of zero local energy and SL passivity can be substantial. We pursue this further in the following section.

The Gibbs states (\ref{eq:stmix}) with population probabilities
(\ref{eq:gib}) of the particle pair ${\cal S}_2$ have partition function
\begin{displaymath}
{\cal Z} = \sum_{j=0}^3 \exp\left( -\frac{E_j}{kT} \right)
= 2\left(\cosh\frac{\kappa}{kT}+\cosh\frac{m}{kT} \right)
\end{displaymath}
\noindent
and, in particular,
\begin{equation}
\delta_0 = \frac{2}{{\cal Z}}\sinh\frac{m}{kT} \; , \;\;\;
\delta_1 = \frac{2}{{\cal Z}}\sinh\frac{\kappa}{kT} \; .
\label{eq:dels}
\end{equation}
The bottom display in Fig.\ 1 uses (\ref{eq:dels}) to show local energy varying by temperature for
$\kappa=2$ through the Gibbs states (red path in the top display), from the $T=0$ ground state at
$(\delta_0,\delta_1)=(1,0)$ to the $T=\infty$ completely mixed state at $(\delta_0,\delta_1)=(0,0)$.
The Gibbs states exit the $\Omega_{\mbox{\tiny o}}=0$ region at a non-zero temperature $T_*$
for any $\kappa>0$, where the critical temperature $T_*$ is determined by the condition
$|\eta \xi| <1-\eta^2$ in (\ref{eq:maxxx}) with (\ref{eq:dels}) used in (\ref{eq:etxi}).

The particle pair ${\cal S}_2$ with Hamiltonian (\ref{eq:ham2}) is a special case of the class
of two-particle systems ${\cal S}_{2,\gamma}$ with Hamiltonian
\begin{equation}
\mbox{\bf H} =
\kappa \left( \frac{1+\gamma}{2} \mbox{\boldmath$\sigma$}_1^x \mbox{\boldmath$\sigma$}_2^x
+ \frac{1-\gamma}{2} \mbox{\boldmath$\sigma$}_1^y \mbox{\boldmath$\sigma$}_2^y \right)
+ \mbox{\boldmath$\sigma$}_1^z
+ \mbox{\boldmath$\sigma$}_2^z
\label{eq:ham2an}
\end{equation}
\noindent
where $\gamma\in[0,1]$ is the coupling anisotropy. The coupling is isotropic when $\gamma=0$ and
fully anisotropic when $\gamma = 1$ as in (\ref{eq:ham2}). (Re)define $m=\sqrt{\gamma^2\kappa^2+4}$
for (\ref{eq:ham2an}). Then the eigenenergies of (\ref{eq:ham2an}) are those in (\ref{eq:en}).
The class ${\cal S}_{2,\gamma}$ of anisotropic systems is interesting because it allows us to see
firsthand the roles of ground state nondegeneracy and entanglement in our theorem: for
$\gamma = 0$ and $\kappa<2$ the ground state $|E_0\rangle$ is nondegenerate but separable, and
for $\gamma\in (0,1)$ it is degenerate if and only if
\begin{equation}
\kappa = \frac{2}{\sqrt{1-\gamma^2}} .
\label{eq:kapcond}
\end{equation}

To determine the local energy in a particle of ${\cal S}_{2,\gamma}$, we again consider a general
local quantum operation $\cal G$ applied to a particle of ${\cal S}_{2,\gamma}$, where $\cal G$ has
Kraus operators (\ref{eq:ko}) with constraints (\ref{eq:krc}). The energy extracted from
${\cal S}_{2,\gamma}$ in an eigenmixture state $\rho$ by the local operation $\cal G$ is, after
some calculation,
\begin{eqnarray}
&& \!\!\!\!\!\!\!\!\!\!\!\!\!\!\!\!
\Delta E(\rho) = (1-\eta) \mbox{\bf u}^\dagger \mbox{\bf u}
- (1+\eta)\mbox{\bf t}^\dagger \mbox{\bf t}                                         \nonumber \\
&& \;\;\; + \, \xi\frac{{\mathbf s}^\dagger {\mathbf v} + {\mathbf v}^\dagger {\mathbf s}}{2}
+ \mu\frac{\mbox{\bf u}^\dagger {\mathbf t} + {\mathbf t}^\dagger \mbox{\bf u}}{2} - \xi
\label{eq:der}
\end{eqnarray}
\noindent
where
\begin{equation}
\eta = \frac{2\delta_0}{m} \, , \;\;
\xi = \frac{\gamma^2\kappa^2\delta_0}{m} + \kappa\delta_1 \, , \;\;
\mu = \frac{\gamma\kappa^2\delta_0}{m} + \gamma\kappa\delta_1     \label{eq:exm}
\end{equation}
\noindent
with $\delta_0 = p_0-p_3$ and $\delta_1 = p_1-p_2$. To maximize (\ref{eq:der}) subject to
(\ref{eq:krc}), we again take essentially the approach in \cite{frey} and find that the
maximum of $\Delta E(\rho)$ in (\ref{eq:der}) subject to constraints (\ref{eq:krc}) is
the unconstrained maximum of
\begin{eqnarray}
&& \!\!\!\!\!\!\!\!\!\!\!\!\!\!\!\!
\omega(\alpha,\beta) = (1-\eta)\sin^2\alpha-(1+\eta)\sin^2\beta    \nonumber \\
&&  \;\;\;\; +\, |\xi|\cos\alpha\cos\beta + |\mu|\sin\alpha\sin\beta - \xi   .  \label{eq:ome3}
\end{eqnarray}
\noindent
The maximum of $\omega(\alpha,\beta)$ is the local energy of an ${\cal S}_{2,\gamma}$ particle.
Using (\ref{eq:dels}) for $\delta_0$ and $\delta_1$ in (\ref{eq:exm}), we numerically maximize
$\omega(\alpha,\beta)$ to find (see Fig.\ 2) the Gibbs states critical temperatures $T_*$ as a
function of the coupling strength $\kappa$ for selected anisotropies $\gamma$. The points in
Fig.\ 2's inset show the values of $\kappa$ where, for different $\gamma\in (0,1)$, $T_*$ falls
to zero. The superposed curve in the inset is condition (\ref{eq:kapcond}) for degeneracy. For
$\gamma\in(0,1)$ we see that, consistent with our theorem, $T_*=0$ wherever the system coupling
parameters combine in (\ref{eq:kapcond}) to make $|E_0\rangle$ degenerate. For $\gamma=0$, $|E_0\rangle$
is nondegenerate and separable for $\kappa <2$, degenerate for $\kappa=2$, and nondegenerate and
entangled for $\kappa >2$, while Fig.\ 2 shows that $T_*>0$ (extant ground state neighborhood of
SL passivity) only for $\kappa>2$. These various cases illustrate the point of our theorem: a
nondegenerate, entangled ground state is sufficient to create a ground state neighborhood of
SL passivity.

\begin{center}
{\bf IV.\ SL PASSIVITY'S EXTENT}
\end{center}
\vspace*{-1mm}

We explore in three directions the extent of SL passivity's presence in finite quantum systems.
We first ask whether the threshold ground state probability $p_\ast$ identified in our theorem applies
also for system states that are not eigenmixtures. We show by an example that $p_\ast$ does not
necessarily apply to system states with some coherence among eigenstates; that is, given some
coherence, a state with ground state population probability $p_0 > p_\ast$ can fail to be SL passive.
This shows that eigenmixtures, which play a distinctive role in state passivity, are also special for
SL passivity. We then go on to ask whether SL passivity is limited to just systems of small dimension.
We show by the example of a Heisenberg chain of $N$ spin-$\frac{1}{2}$ particles that SL passivity
can be a nonvanishing feature of a many-particle system. Finally, we ask whether SL passivity
is always confined just to eigenmixtures near the ground state. We saw in the previous section
that for the two-particle system ${\cal S}_2$ any eigenmixture with large enough ground state
population probability ($p_0>.9383$ for $\kappa=2$) is SL passive. In this section we offer an
example of a two-particle system in which the ground state's SL passivity neighbourhood extends
all the way to the completely mixed state ($p_0=.25$), and in which, in particular, Gibbs states of
{\it any} temperature are SL passive. Our point with this section's examples is that eigenmixtures are
central to SL passivity and that, among eigenmixtures, SL passivity is neither limited to just quantum
systems of a few particles, nor is it necessarily confined to just the near vicinity of the ground state.

\begin{figure}[b!]
\centerline{\scalebox{1}{\includegraphics{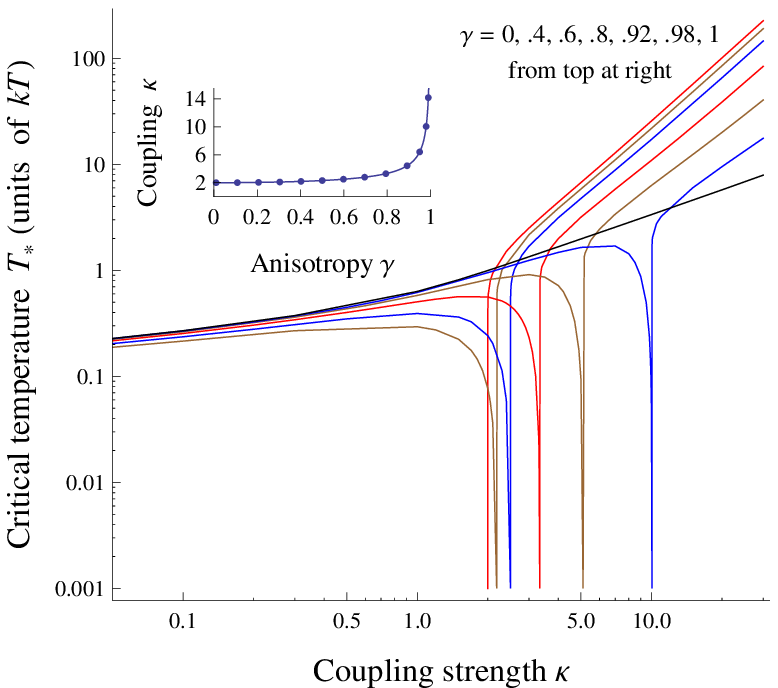}}}
\vspace*{4mm}
\begin{center}
\parbox{3in}{\small
FIG.\ 2. Critical temperatures $T_*$ below which $\Omega_\circ = 0$ for selected coupling
anisotropies $\gamma$. The points in the inset are coupling strengths $\kappa$ where $T_* = 0$.
The superposed curve in the inset is condition (\ref{eq:kapcond}) for ground state degeneracy.}
\end{center}
\end{figure}

Suppose the state $\rho$ of the particle pair ${\cal S}_2$ of the previous section is an eigenmixture.
The local energy of a particle is then $\Omega_\circ=\Omega_\circ(\eta,\xi)$ in (\ref{eq:maxxx}),
and a threshold ground state population probability $p_\ast <1$ exists such that $\rho$ with
$p_0 \geq p_\ast$ is SL passive. Now introduce some coherence to the eigenmixture $\rho$
and consider the system state
\begin{equation}
\rho^\prime = \rho + r (|E_2\rangle\langle E_0| + |E_0\rangle\langle E_2| )
\label{eq:rhoexpr}
\end{equation}
\noindent
with real coherence $r$ where $|r|\leq\sqrt{p_0p_2}$. We find after some calculation
that, for ${\cal S}_2$ in the state $\rho^\prime$, the energy extracted by a general
quantum operation ${\cal G}$ applied locally to a particle is
\begin{eqnarray}
&& \!\!\!\!\!\!\!\!\!\!\!\!\!\! \Delta E(\rho^\prime) = \Omega_\circ(\eta,\xi) + \frac{r}{\sqrt{m(m+2)}}
\left[ \kappa (\mbox{\bf t}^\dagger \mbox{\bf s} + \mbox{\bf s}^\dagger \mbox{\bf t}) \right. \nonumber \\
&& \;\;\;\;+ \kappa^2 (\mbox{\bf u}^\dagger \mbox{\bf s} + \mbox{\bf s}^\dagger \mbox{\bf u}) 
+ (m+2) (\mbox{\bf u}^\dagger \mbox{\bf v} + \mbox{\bf v}^\dagger \mbox{\bf u}) \nonumber \\
&& \;\;\;\, \left. 
- \kappa (m+2) (\mbox{\bf t}^\dagger \mbox{\bf v} + \mbox{\bf v}^\dagger \mbox{\bf t})
\right] \, .
\label{eq:delE}
\end{eqnarray}
\noindent
Suppose ${\cal G}$ has a single Kraus operator
$\mbox{\bf K} = \exp(-i\phi \mbox{\boldmath$\sigma$}^y)$.
The energy (\ref{eq:delE}) extracted by this (unitary) ${\cal G}$ is
\begin{equation}
\Delta E(\rho^\prime ) = \frac{2 \sin^2 \phi}{m\kappa} \left[ r A \cot \phi
- \eta -2\xi \right]
\end{equation}
\noindent
where
\begin{displaymath}
A=\frac{2}{\kappa}\sqrt{\frac{m-2}{m}}(2+(m+\kappa)(\kappa+1)) \, .
\end{displaymath}
\noindent
For ${\cal S}_2$ with any degree of coupling $\kappa$ and any $\rho^\prime$ in (\ref{eq:rhoexpr})
with nonzero coherence $r$, we can pick the angle $\phi$ associated with ${\cal G}$ so that $\Delta E(\rho^\prime )$
is positive. This is an example in which the smallest amount of coherence added to a SL passive eigenmixture $\rho$
renders the resulting state $\rho^\prime$ not SL passive, allowing energy to be extracted from the system.
Eigenmixtures play a distinctive role in state passivity; this shows that they do also in SL passivity.

The particle pair ${\cal S}_2$ is a case of an $N$-particle closed Heisenberg spin chain
${\cal S}_N$ with Hamiltonian
\begin{equation}
\mbox{\bf H} = \kappa \prod_{i=1}^N \mbox{\boldmath$\sigma$}_i^x \mbox{\boldmath$\sigma$}_{i+1}^x
+ \sum_{i=1}^N \mbox{\boldmath$\sigma$}_i^z
\label{eq:hamNr}
\end{equation}
\noindent
where $\mbox{\boldmath$\sigma$}_{N+1}^x \equiv \mbox{\boldmath$\sigma$}_1^x$. For each $N \geq 2$
the ground state $|E_0\rangle$ is nondegenerate and fully entangled. Suppose ${\cal S}_N$ is in
state (\ref{eq:stmix}) with $d=2^N$ and we apply a general quantum operation $\cal G$ to a particle.
As in the two-particle case, $\cal G$ involves at most four Kraus operators (\ref{eq:ko}) with
the constraints (\ref{eq:krc}). We seek the system energy $\Delta E(\rho)$ in (\ref{eq:sysen})
extracted by $\cal G$ from ${\cal S}_N$ and find, remarkably, that the extracted energy
$\Delta E(\rho)$ has the same form (\ref{eq:Ome}) for all $N \geq 2$, where $\eta$ and $\xi$ vary
according to $N$. Therefore, for $N\geq 2$ the local energy in a particle is
(\ref{eq:maxxx}) with $\eta$ and $\xi$ depending on $N$. We suppose $\rho$ is a Gibbs state for each
$N$ and then solve $|\xi\eta|=1-\eta^2$ in (\ref{eq:maxxx}) to obtain the critical temperature
$T_*$ for zero local energy. Figure 3 shows the results of these calculations for spin chains
${\cal S}_N$ of up to six particles. (For $N\geq 6$ the curves for $T_*$ are visually
indistinguishable.) We see that $T_*$ increases with $N$, but that that increase quickly becomes
vanishingly small. This could be expected for a closed chain's ring topology. A particle is most
strongly effected by its two immediate neighbors, while any added particle joins the chain as a
most distant particle. Increasing $N$ only adds distant neighbors with vanishingly less effect,
and Fig.\ 3 reflects this. Most importantly, Fig.\ 3 shows that SL passivity and zero local
energy are not limited to just a few particles; theoretically, a neighborhood of SL passivity
can exist with no diminution in systems of arbitrarily many particles.

\begin{figure}[t!]
\centerline{\scalebox{1}{\includegraphics{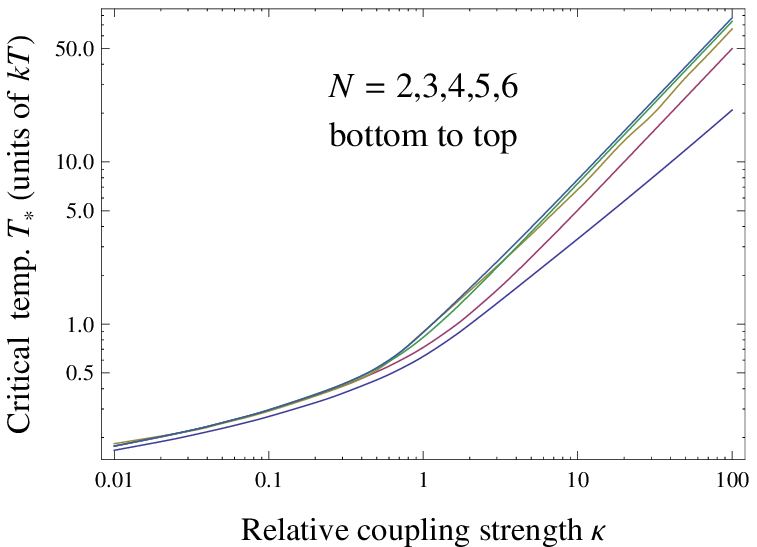}}}
\vspace*{2mm}
\begin{center}
\parbox{2.9in}{\small
FIG.\ 3. Gibbs states' critical temperatures $T_*$ for SL passivity and zero local energy
in $N$-particle spin chains. Curves for $N=6$ and beyond are visually indistinguishable.}
\end{center}
\vspace{-.2in}
\end{figure}

Thus far the spin systems in our examples have all exhibited a SL passivity neighborhood of bounded
extent with, specifically, a finite critical Gibbs temperature $T_* < \infty$. Now consider the
system ${\cal S}_X$ of two Heisenberg XXX-coupled spin particles with Hamiltonian
\begin{equation}
\mbox{\bf H}_X = \mbox{\boldmath$\sigma$}_1^x \mbox{\boldmath$\sigma$}_2^x
+ \mbox{\boldmath$\sigma$}_1^y \mbox{\boldmath$\sigma$}_2^y
+ \mbox{\boldmath$\sigma$}_1^z \mbox{\boldmath$\sigma$}_2^z    \, ,
\label{eq:hamX}
\end{equation}
\noindent
eigenenergies $E_0=-3$, $E_1=E_2=E_3=1$, and corresponding eigenstates
\begin{eqnarray}
&& \!\!\!\!\!\! |E_0\rangle = \frac{|10\rangle-|01\rangle}{\sqrt{2}} , \;\;
|E_2\rangle = |00\rangle , \;\;   \nonumber     \\
&& \!\!\!\!\!\! |E_1\rangle = \frac{|10\rangle+|01\rangle}{\sqrt{2}} , \;\;
|E_3\rangle = |11\rangle .  \nonumber
\end{eqnarray}
\noindent
The ground state $|E_0\rangle$ of ${\cal S}_X$ is nondegenerate and entangled so we conclude
that $|E_0\rangle$ has a neighborhood of SL passivity. To determine this neighborhood's extent,
we derive for ${\cal S}_X$ the energy (\ref{eq:sysen}) extracted by a general quantum operation
(\ref{eq:gqop}) on one of ${\cal S}_X$'s two particles, finding that
\begin{eqnarray}
&& \!\!\!\!\!\!\!\!\!\!\!\!\!\!\!\!\!\!\!
\Delta E(\rho) = -(p_0+p_1-2p_2) {\mathbf u}^\dagger  {\mathbf u}              \nonumber \\
&& \;\;\;\;\;\;\;\; -(p_0+p_1-2p_3){\mathbf t}^\dagger  {\mathbf t}              \label{eq:sX} \\
&& \;\;\;\;\;\;\;\; - (p_0-p_1)(2-{\mathbf s}^\dagger  {\mathbf v} -{\mathbf v}^\dagger  {\mathbf s})  \nonumber
\end{eqnarray}
\noindent
for any eigenmixture $\rho$ with population probabilities $p_0,p_1,p_2,p_3$. For Gibbs states
$p_0\geq p_1\geq p_2\geq p_3$. Also, ${\mathbf u}^\dagger  {\mathbf u}\geq 0$, ${\mathbf t}^\dagger {\mathbf t}\geq 0$,
and ${\mathbf s}^\dagger  {\mathbf v} +{\mathbf v}^\dagger  {\mathbf s} \leq
{\mathbf s}^\dagger  {\mathbf s} +{\mathbf v}^\dagger  {\mathbf v}\leq2$. We readily conclude then
that $\Delta E(\rho)\leq 0$ in (\ref{eq:sX}), and that the local energy in a particle of ${\cal S}_X$
is $\Omega_\circ = 0$ for any Gibbs state $\rho$; that is, $T_* = \infty$. Thus ${\cal S}_X$ is a
quantum system whose Gibbs states of all temperatures are both passive and SL passive.
 
\begin{center}
{\bf V. SUMMARY AND FINAL REMARKS}
\end{center}
\vspace*{-1mm}

We summarize the work presented in this paper by emphasizing that SL passivity provides a
framework for determining the energy that is locally accessible in multipartite quantum systems.
This newly identified property of states in finite quantum systems is a variant of the standard
notion of state passivity, where the nature of the operation on the multipartite system is both 1) relaxed
to allow any general quantum operation and 2) restricted in its application to a subsystem. These
countervailing modifications yield unexpected and interesting results. Passive states are known to
be eigenmixtures that have no population probability inversion and, in particular, all Gibbs states are
passive. While eigenmixtures are similarly important to SL passivity, the conditions for and the extent
of SL passivity within these states are more subtle. If the ground state is nondegenerate and
entangled, then the system exhibits a neighborhood of SL passivity around the ground state. Using
Gibbs state temperature to gauge this neighborhood's extent, we saw by example that the Gibbs state
critical temperature for SL passivity can be $T_*=0$ (when the ground state is separable or degenerate),
positive and finite, or even $T_* = \infty$ (in which case all the Gibbs states are SL passive). The
existence of systems with $T_*=\infty$ decisively establishes that SL passivity is not limited to just
the near vicinity of the ground state; its extent can be considerable. Remarkably, SL passivity can
extend, also, without diminution to high-dimensional systems of arbitrarily many particles. The Gibbs critical
temperature of an $N$-particle Heisenberg ring, for example, quickly converges for increasing $N$ to
a positive limit value $T_*>0$. For such a system in a state of SL passivity, the system particles act
collectively to block energy release from any one particle.

Our theorem concerns energy extracted by a local operation when the system state is near the
ground state. A complementary result can be stated for {\em adding} energy when the system state
is near the {\em maximum} energy eigenstate $|E_{d-1}\rangle$. Let $\cal S$ be a finite quantum system
with Hamiltonian (\ref{eq:spec}) and a specified subsystem $\cal C$. Suppose that $|E_{d-1}\rangle$
is nondegenerate and that in $|E_{d-1}\rangle$, $\cal C$ is fully entangled with the rest
of $\cal S$. Then a threshold maximum energy state population probability $q_*<1$ exists
such that no energy can be added to the system by any local quantum operation on ${\cal C}$
when the system state is an eigenmixture (\ref{eq:stmix}) with $p_{d-1} \geq q_*$. The
two threshold probabilities $p_\ast$ and $q_\ast$ associated with a subsystem ${\cal C}$
are not generally equal. The proof of this complementary result parallels that of our theorem.
 
Strong local passivity is just newly discovered, and it is premature to anticipate applications.
We note, though, that in the anisotropically coupled particle-pair ${\cal S}_{2,\gamma}$, the
critical Gibbs temperature $T_*$ is highly sensitive to the strength of the external magnetic field
(reflected in the parameter $\kappa$) when condition (\ref{eq:kapcond}) is close to satisfied.
In fact, under conditions close to (\ref{eq:kapcond}), Fig.\ 2 shows $T_*$ varying over orders of
magnitude in response to only small change in $\kappa$. The critical temperature $T_*$ is a proxy
for the extent of the SL passivity neighborhood, and with a suitable initial system state, varying
$T_*$ can switch on/off the SL passivity of a subsystem, locking up or allowing the release of
energy. This suggests that a system such as ${\cal S}_{2,\gamma}$ might be a sensitive detector
of small changes in the external magnetic field, or by actively modulating the external field,
${\cal S}_{2,\gamma}$ might be used as a switch for energy release. These comments, while only
speculative, suggest potential possibilities.

The notion of SL passivity raises a host of theoretical questions. In particular, SL passivity makes a
new connection between the local/global paradigm in quantum information science and the standard
notion of passivity in thermodynamics, potentially advancing, for example, the theory of
quantum Maxwell demons for subsystems.

\begin{center}
{\bf ACKNOWLEDGMENT}
\end{center}
\vspace{-1mm}

Fruitful conversations with Yu Watanabe and Jos\'e Trevison are gratefully acknowledged. K.F.
acknowledges support from JSPS (grant no.\ 254105).

\setlength{\baselineskip}{4.2mm}

\end{document}